\documentclass[12pt]{article}
\usepackage{pdproc,epsfig} 

\usepackage{amsmath}

\begin{document}


\renewcommand{\thefootnote}{\alph{footnote}}
 
\title{STATUS OF THE DOUBLE CHOOZ EXPERIMENT}

 \author{J. V. DAWSON on behalf of the Double Chooz collaboration}

\address{Laboratoire Astroparticule et Cosmologie,\\ 10 rue Alice Domon et L\'eonie Duquet,\\ 75205 Paris, France \\{\rm E-mail: jdawson@in2p3.fr}}

%\author{}

%\address{}

\abstract{The Double Chooz experiment is the first of the next
  wave of reactor experiments searching for a
  non-vanishing value of the mixing angle $\theta_{13}$. The
  experimental concept and detector design are presented, and the most pertinent
  backgrounds are discussed. Operation of
  the far detector is expected to begin by the end of
  2009. Installation of the near detector will occur in 2010.  Double Chooz
  has the capacity to measure sin$^2(2\theta_{13})$ to 3$\sigma$ if
  sin$^2(2\theta_{13})>0.05$ or exclude sin$^2(2\theta_{13})$ down to
  0.03 at 90\% for $\Delta m_{31}^2 = 2.5 \times 10^{-3} eV^2$ with three
  years of data with both near and far detectors. 
  }

\normalsize\baselineskip=15pt

%\begin{keyword}

%\PACS 
%\end{keyword}

% main text
\section{Introduction}
\label{intro}
Neutrino oscillation has been clearly established via the study of
solar, astmospheric, reactor and beam neutrinos.  Combination of these
results requires the existence of (at least) three-neutrino mixing. In
the current view, the PMNS mixing matrix relates the three neutrino
mass eigenstates to the three neutrino flavour eigenstates
parameterized by three mixing angles ($\theta_{12}$, $\theta_{13}$ and
$\theta_{23}$) and one CP violating phase $\delta_{cp}$ (for Dirac
neutrinos).  Great progress has been made in measuring the mixing
angles and the two squared  mass differences $\Delta m^2_{ij} = m^2_i - m^2_j$ (for a good review of neutrino oscillation experiments
see for example \cite{Dore2008}). However, the mixing angle $\theta_{13}$, the mass
hierarchy and the $\delta_{cp}$ phase are  still currently
unknown. Indeed only upper limits to the value of  $\theta_{13}$ have
been found, indicating that this angle is very small with respect to
the other two.  Whilst a measurement of  $\theta_{13}$ would complete
the knowledge of the mixing angles, even a more stringent upper limit would
be useful since the size of $\theta_{13}$  has a great bearing on the
possibility to observe CP violation in the leptonic sector with
upcoming neutrino experiments (see for example \cite{Schwetz2007} for
a discussion of  $\theta_{13}$ and CP violation discovery in
forthcoming experiments).

A three-flavour global analysis on existing data gives an upper bound
of sin$^2\theta_{13} < 0.035$ at 90\% C.L \cite{Schwetz2008}. This
value is dominated by the bound given by the reactor experiment, CHOOZ
\cite{apollonio2003}, in which no oscillation was observed $R = 1.01
\pm 2.8\%(stat) \pm 2.7\%(sys)$.  

%reactor experiments
Reactor experiments search for the disappearance of electron anti-neutrinos emitted from the cores of the nuclear reactors.  Equation \ref{eqn:survival} gives the survival probability of a $\bar{\nu_e}$ from a reactor, where $E$ is the neutrino energy and $L$ is the distance from the source to the detector.

%put in survival probability..
\begin{equation}
\begin{split}
 P(\bar{\nu_e}\rightarrow \bar{\nu_e}) = 1 - sin^2(2\theta_{13}) sin^2\frac{\Delta m_{31}^2 L}{4E} - cos^4 \theta_{13} sin^2(2\theta_{12})sin^2\frac{\Delta m_{21}^2 L}{4E}\\ + 2sin^2\theta_{13} cos^2\theta_{13} sin^2\theta_{12}\left(cos \frac{(\Delta m_{31}^2 - \Delta m_{21}^2)L}{2E} - cos \frac{\Delta m_{31}^2 L}{2E} \right)
\end{split}
\label{eqn:survival}
\end{equation} 
For short baselines only the first two terms are relevant.
With a well positioned detector (such that L/E is $\sim$0.3 km/MeV), a
detector might observe less neutrinos than anticipated indicating a
non-zero value of $\theta_{13}$ and therefore these experiments are
termed 'disappearance' experiments. 'Appearance' experiments i.e. long
baseline accelerator experiments aim to measure the appearance of $\nu_e$s in a
$\nu_\mu$ beam. 

Reactor based $\theta_{13}$ experiments have some advantages over long baseline
accelerator experiments.  They suffer less from parameter
degeneracies, being independent of $\delta_{cp}$ and the sign of
$\Delta m_{31}$ and having only a weak dependence of $\Delta m^2_{21}$. Since the neutrino
energies are low, $\sim$1 to 10 MeV, and the detectors are positioned at
short distances, there are no matter effects.  The major drawback to
this type of experiment is that there is limited knowledge on the
neutrino production processes inside the reactors. 

%Double Chooz concept
\section{Double Chooz}
The Double Chooz experiment \cite{LOI2006} is located at Chooz, the same site as the
original Chooz experiment, in the Champagne-Ardennes region in France.
The site contains two closely neighbouring nuclear reactors each with
a thermal power of 4.27 GW.  The Double Chooz concept is to use two identical
detectors; one near, to effectively measure the neutrino spectrum and
flux from the reactor, and one far, to observe any neutrino disappearance.

The far detector is located in the same underground laboratory as the
original Chooz experiment (1 km from the two cores). This site is
perfect for three reasons; an ideal L/E of 0.3 MeV/km, the cost is significantly reduced due to the existing laboratory, and
the experimental background rate i.e. from muons, neutrons and rock
radioactivity etc are already well measured with reactor-off data.  The near detector
underground laboratory will be 400m from the two reactors and must be
constructed.

The target is a Gadolinium loaded scintillator, with an interacting
anti-neutrino of energy greater than 1.8 MeV causing an inverse-beta decay of a proton.
\begin{equation}
\bar\nu_e + p \rightarrow n + e^+ 
\end{equation}
The positron slows depositing its kinetic energy in the scintillator. It
quickly annihilates; releasing two 511 keV gammas.  The total prompt
visible energy seen is some 1 to $\sim$8 MeV and is directly
related to the energy of the neutrino $E_\nu = E_{visible} + 0.8$MeV.
After a characteristic delay, the neutron slows and is captured; on
Gadolinium (absorption time of 30 $\mu$s) or on Hydrogen.  Gamma cascades from the captures give energy
deposits of $\sim$8 MeV (from Gadolinium) and 2.2 MeV from Hydrogen.

As the interaction cross-section rises (with the square of the
energy) and the reactor neutrino spectrum falls in a similar fashion,
the convolution of these two, the observed spectrum is roughly
Gaussian in shape with a peak visible energy of $\sim$4~MeV.

\section{Detector Design}

\begin{figure}[ht]
%\vspace*{13pt}
\begin{center}
\leftline{\hfill\vbox{\hrule width 10cm height0.001pt}\hfill} 
\vspace*{10pt}
\mbox{\epsfig{figure=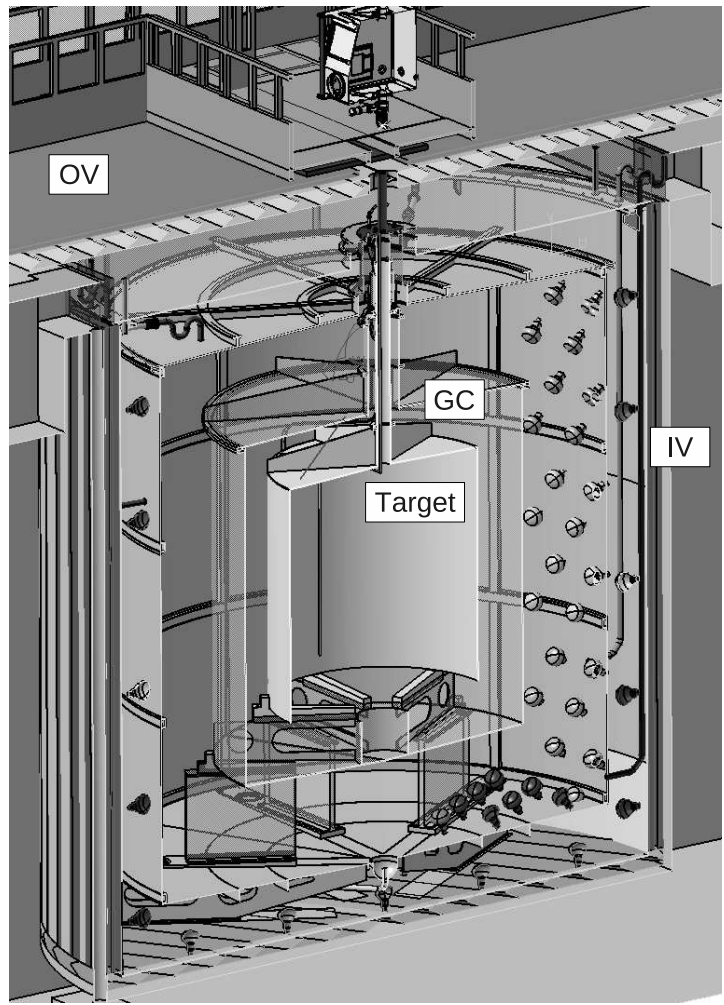,width=10.0cm}}
\leftline{\hfill\vbox{\hrule width 10cm height0.001pt}\hfill}
%\vspace*{1.4truein}		%ORIGINAL SIZE=1.6TRUEIN x 100% - 0.2TRUEIN
\caption{Design of detector. OV is Outer Veto, IV is Inner Veto, GC is Gamma-Catcher.}
\label{fig:detector}
\end{center}
\end{figure}

Figure \ref{fig:detector} shows the detector and laboratory
design. Both detectors are identical from the buffer
tank (inner-most stainless steel vessel) inwards which is a physics
requirement. Shielding against the radioactivity of the rock is
provided by 15 cm of demagnetised steel for the far detector but less
stringent shielding is required for the near detector.

Each detector is formed from a series of nested cylinders with each
volume filled with different liquids; insensitive buffer oil for
shielding, Gd-doped scintillator as the target and undoped
scintillators for gamma rays, fast neutrons and muons.

The two inner vessels are acrylic and transparent to photons
above 400 nm. The inner-most vessel is the Target, with a diameter of 2.3 m, which contains 10 m$^{3}$ of Gadolinium-doped scintillator; such that the
scintillator contains 1 g/l of Gadolinium. In this volume
neutron-capture on Gadolinium can occur releasing cascade gammas with
an energy of $\sim$8 MeV.  More than 80\% of neutron captures are on
Gadolinium rather than Hydrogen. The definition of a neutrino candidate
event is one in which neutron capture on Gadolinium occurs.

Enclosing the target is the Gamma-Catcher volume, with a diameter of
3.4m, which contains 22 m$^{3}$ of
undoped scintillator. The purpose of this volume is to detect the
gammas emitted in both the neutron-capture process and positron annihilation in the target, such that
gammas emitted from neutrino events occurring in the outer volume of the
target are detected. This results in a well-defined target volume. 

Since the photomultipliers are the most radioactive component of the
detector, the inner volumes are shielded by a buffer volume, with a
diameter of 5.5m, filled with non-scintillating paraffin oil.  Events
occurring in the acrylic volumes are detected by 390 10 inch low background photomultiplier
tubes (Hamamatsu R7081 \cite{hamamatsu}) fixed to the inside of the steel buffer tank.  Uniquely the photomultiplier tubes are angled
to improve the uniformity of light collection efficiency in the
inner-most volumes.  We anticipate to achieve 7\% energy resolution at 1 MeV.  

The outer detector volume is steel walled, with a diameter of 6.6m,
and filled with scintillator. 78 8 inch photomultipliers (Hamamatsu R1408\cite{hamamatsu}) line the outermost wall which is painted with a reflective
white coating. This volume is the Inner Veto
with the purpose of detecting and tracking muons and fast neutrons.

On top of the detector sits
the Outer Veto. This comprises strips of plastic scintillator and
wavelength-shifting fibres. The veto extends further than the detector
diameter with the purpose of detecting and tracking muons.  The
precision of the entry point of a muon, X-Y position, will be far
more precise than that achieved by the Inner Veto and detector.  One
of the main objectives is to tag near-miss muons, which interact in the
surrounding rock (and not in the detector) but produce fast
neutrons. Another important goal is to determine whether a muon
entered the inner detector. Muons that do so can produce cosmogenic
isotopes (i.e. via a photonuclear interaction on $^{12}$C), some of
which will produce backgrounds for the experiment.

%change picture...

\section{Backgrounds}\label{backgrounds}
As each neutrino produces  two time-correlated signals; that of
the positron and a delayed capture of a neutron (with characteristic
decay time of 30 $\mu$s), backgrounds can come from two sources;
accidental and time-correlated.

The accidental component comes from the random chance that two events
of appropriate energy interact within this characteristic time. Since
these two events are unrelated this rate can easily be measured, based
on the singles rate.  The main source of events come from
radioactive contamination with the dominant source being the
photomultiplier tubes. For the accidental component to be well
constrained, strict radioactive contamination limits have been placed
on all parts.

The most difficult backgrounds to study are those that are, like our
signal, time-correlated. From the experience of Chooz it is
anticipated that Double Chooz will observe some $\sim$ 1.5 events/day
of false neutrino-like events. The Chooz experiment had a period of
data-taking before operation of the nuclear reactors began and so the
background could be very thoroughly investigated.  The sources of the
neutrino-like events observed were attributed to fast neutrons
(muon-induced neutrons) and
cosmogenically produced isotopes (also muon produced).

Fast neutrons can mimic neutrino signals by producing a
proton-recoil (positron-like signal) and a delayed neutron
capture. If the muon is seen by the experiment then these events can be
tagged. More dangerous, however, are near-miss muons which
interact in the rock releasing fast neutrons which interact in the
detector.  The primary purpose of the Outer Veto is to identify these
events by covering an area wider than the detector itself.

Those cosmogenically produced isotopes that are dangerous for the
experiment are those that result in electron emission followed by neutron
emission, as these mimic well our neutrino signal. Two isotopes,
$^{8}$He and $^{9}$Li, have long decay times (119 ms and 174 ms
respectively)\cite{Hagner2000} rendering a hardware veto impractical.  Coupling
information from the Outer Veto (with precise muon entry points),
Inner Veto and inner detector will allow reconstruction of muon
tracks to identify muons that cross the inner detector.

\section{Improvements on Chooz}\label{improvements}
Improvements on the original Chooz experiment have been made in two
ways; the detector design and the two-detector concept.
The new detector target is more than twice as large as the original
Chooz detector.  The scintillator technology has improved, and
Gadolinium-loaded scintillator now is
very stable (on the timescale of years) allowing a longer run time
$\sim$5 years.  The number of neutrinos detected in the far detector
assuming 3 years of running will be $\sim$60,000 compared to 2,700 in
the Chooz experiment, reducing the statistical error, 2.8\% in Chooz,
to 0.4\%.

The aim is to reduce the systematic error, 2.7\% in Chooz, to less
than 0.6\%.  There are three sources of systematic error; the reactor,
the detector and the analysis. With two detectors, each reactor
component systematic; flux and cross-section, reactor power and
energy per fission, reduce to below 0.1\%.  Making a relative
measurement, between the two detectors, reduces many of detector
systematics to similar orders.

The scintillators will be produced for both detectors in one batch,
reducing the systematic on the number of H and Gd atoms in each
detector.  With a well performing scintillator the number of observed photons should be high enough such that all of the positron signal is observed so there
is no systematic introduced by cutting on the positron spectrum. The improved detector design, with target and gamma-catcher vessels, provides a fixed fiducial volume such that positional cuts are not needed in the analysis eliminating another important source of systematic error.

In general, controlling the relative systematics between the two
detectors is far easier than the absolute.  Two detectors, however,
introduces one new systematic - the live time, as both detectors must
operate simultaneously.

\section{Construction Progress} \label{progress}
The near detector site has been chosen, some $\sim$400 m from the two
reactors, and the civil engineering study made.  The excavation and
construction of the new laboratory is foreseen to be completed by the
end of 2010.  The new laboratory will be slightly deeper than the site
originally proposed giving a shielding of 115
m.w.e (metres water equivalent). At this new site we anticipate
detecting $\sim$ 500 neutrinos per day.  

Good progress has been made on the far detector and related
infra-structure.  The far detector shielding, buffer vessel and inner
veto photomultipliers have been installed. The laboratory has been
thoroughly cleaned, all surfaces painted and clean tents and a
protective wall installed. During March 2009, the buffer vessel was simultaneously welded and lowered in to place. Scaffolding was 
erected inside this vessel so that the inner detector photomultipliers
can be installed. After this is complete, the inner acrylic vessels
can be installed.
The liquid handling systems will be completed in parallel. Already the
scintillator (and oil) tanks and associated filling systems have been
installed in the liquid storage facility close to the entrance of the
tunnel.
An aggressive schedule is proposed such that the detector is
anticipated to commence operation by the end of 2009.

\section{Conclusion}\label{conclusion}
\begin{figure}
%\vspace*{13pt}
\begin{center}
\leftline{\hfill\vbox{\hrule width 10 cm height0.001pt}\hfill}
\vspace*{10pt}
\mbox{\epsfig{figure=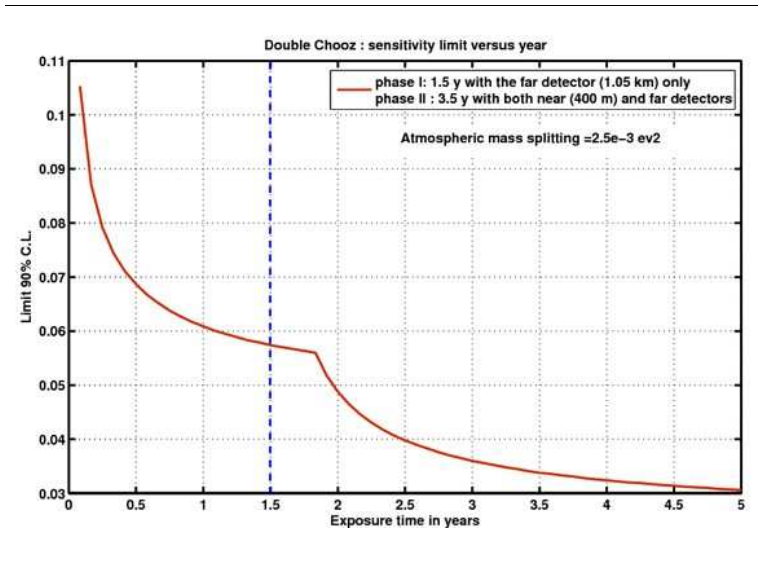,width=10.0cm}}
%\vspace*{1.4truein}		%ORIGINAL SIZE=1.6TRUEIN x 100% - 0.2TRUEIN
\leftline{\hfill\vbox{\hrule width 10 cm height0.001pt}\hfill}
\caption{Sensitivity of the experiment, showing the first
  far detector only phase and the rapid improvement as the near
  detector is included.}
\label{fig:sensitivity}
\end{center}
\end{figure}

Double Chooz will be the first next generation reactor experiment to
commence operation.  The construction of the far detector of the Double Chooz experiment
will be completed in 2009 with detector commissioning occurring at the
end of the year. The first phase of data-taking will occur with the
far detector only.  Figure \ref{fig:sensitivity} shows the improvement
in sensitivity as a function of time; the far detector only phase and
the two detector phase.  Whilst the experiment is less sensitive without the
near detector, it will still be more sensitive than the original Chooz
detector and should reach a sensitivity to sin$^2(2\theta_{13})$ of
0.06 with one year of data. 
With two detectors,  Double Chooz will be able to measure
sin$^2(2\theta_{13})$ to 3$\sigma$ if sin$^2(2\theta_{13})>0.05$ or
exclude sin$^2(2\theta_{13})$ down to 0.03 at 90\% for $\Delta
m_{31}^2 = 2.5 \times 10^{-3} eV^2$ with three years of data with both
near and far detectors.

%%%%%%%%%%%%%%%%%%%%%%%%%%%%%%%%%%%%%%%%%%%%%%%%%%%%%5
  
%\section{Acknowledgements}
%  Acknowledgements should appear just before the references.

\end{document}